\documentclass[conference]{IEEEtran}

\IEEEoverridecommandlockouts

\makeatletter
\def\markboth#1#2{\def\leftmark{\@IEEEcompsoconly{\sffamily}\MakeUppercase{\protect#1}}%
\def\rightmark{\@IEEEcompsoconly{\sffamily}\MakeUppercase{\protect#2}}}
\makeatother

\usepackage[english]{babel}
\usepackage{color}
\usepackage{lipsum}
\usepackage{caption}
\usepackage{cite}
\usepackage[pdftex]{graphicx}
\usepackage{color}
\usepackage{subcaption}
\usepackage{amsmath}
\usepackage{amsfonts}
\usepackage{amssymb}
\usepackage{amsthm}
\usepackage{array}
\usepackage{verbatim}
\usepackage{listings}
\usepackage{algorithm}
\usepackage{algpseudocode}
\usepackage{url}
\usepackage{enumerate}
\usepackage{multirow}
\usepackage{authblk}
\usepackage{epsfig}
\usepackage{epstopdf}
\usepackage{multicol}
\usepackage[font=footnotesize]{caption}
\usepackage{epstopdf}
\usepackage{soul}

\newcommand{\bi}{\begin{itemize}}
\newcommand{\ei}{\end{itemize}}
\newcommand{\be}{\begin{equation}}
\newcommand{\ee}{\end{equation}}

\def\beq{\begin{equation}}
\def\eeq{\end{equation}}
\def\beqa{\begin{eqnarray}}
\def\eeqa{\end{eqnarray}}
\def\beqan{\begin{eqnarray*}}
\def\eeqan{\end{eqnarray*}}

\begin{document}

\title{\mbox{Transport Layer Performance in 5G mmWave Cellular}}
\author[1]{\bf Menglei Zhang}
\author[1]{\bf Marco Mezzavilla}
\author[1]{\bf Russell Ford}
\author[1]{\bf Sundeep Rangan}
\author[1]{\bf Shivendra Panwar}
\author[2]{\\ \bf Evangelos Mellios}
\author[2]{\bf Di Kong}
\author[2]{\bf Andrew Nix}
\author[3]{\bf Michele Zorzi}
\affil[1]{NYU Tandon School of Engineering, USA $^2$University of Bristol, UK $^3$University of Padova, Italy}

%

\maketitle

\begin{abstract}
The millimeter wave (mmWave) bands are likely to play a significant role
 in next generation cellular systems due to the possibility of very high
 throughput thanks to the availability of massive bandwidth and high-dimensional antennas.
Especially in Non-Line-of-Sight conditions, significant variations in the received RF power can occur as a result of the scattering from nearby building and terrain surfaces. Scattering objects come and go as the user moves through the local environment. At the higher end of the mmWave band, rough surface scatter generates cluster-based small-scale fading, where signal levels can vary by more than $20$ dB over just a few wavelengths. This high level of channel variability may present
significant challenges for congestion control.
Using our recently developed end-to-end mmWave ns3-based framework, this paper presents the first performance evaluation of TCP congestion control in next-generation mmWave networks. Importantly, the framework can incorporate detailed models
of the mmWave channel, beamforming and tracking algorithms, and builds on statistical channel models
derived from real measurements in New York City, as well as detailed ray traces.
\end{abstract}


\section{Introduction}

The millimeter wave (mmWave) bands -- roughly corresponding to
frequencies above 10~GHz --  have attracted considerable attention for
next-generation cellular wireless systems
~\cite{KhanPi:11-CommMag,rappaportmillimeter,RanRapE:14,andrews2014will,ghosh2014millimeter}.
Due to the propagation properties at these frequencies,
the use of mmWave bands could create networks
with two features that have never been seen together before, namely
links with \emph{massive peak capacity},
but capacity that is \emph{highly variable}.

The massive peak rates
arise from the tremendous amount of spectrum available in the mmWave bands
-- up to 200 times by some estimates \cite{KhanPi:11-CommMag}.
This very large bandwidth can be
combined with the large number of spatial degrees of freedom available in
high-dimensional antenna arrays.  Indeed, recent prototypes have
demonstrated multi-Gbps throughput in outdoor environments \cite{gozalvez20155g}.
Simulation and analytical studies \cite{AkdenizCapacity:14,BaiHeath:14,nix1}
have also predicted capacity gains that are orders of magnitude
greater than in current cellular systems.

At the same time, mmWave links are likely to have highly variable quality.
MmWave signals are completely blocked by many common
building materials such as brick and mortar,
\cite{Allen:94,Anderson04,Alejos:08,singh2007millimeter,KhanPi:11-CommMag,Rappaport:28NYCPenetratioNLOSs},
and even the human body can cause
up to 35~dB of attenuation \cite{LuSCP:12}.
As a result, the movement of obstacles and reflectors,
or even changes in the orientation of a handset relative to the body or a hand,
can cause the channel to rapidly appear or disappear.

This combination of features -- massive, but highly variable, bandwidth --
has widespread implications for the design of next-generation
networks seeking to exploit these
bands.  The goal of this work is to understand the effect of this high variability introduced by mmWave links
on the \emph{transport layer}, specifically congestion control, an area that to date has been hardly addressed in the literature, to the best of our knowledge.

The fundamental role of congestion control is to regulate
the rate at which source packets
are injected into the network to balance two competing
objectives:  (1) to ensure that sufficient packets are sent to
utilize the available bandwidth, but (2) to avoid too many packets
that may result in congestion to other flows in the network.
From this perspective, mmWave networks -- that may introduce
links with extremely variable
capacity -- raise several key questions.
Will current congestion control mechanisms, e.g., TCP Cubic \cite{levasseur2014tcp},
be effective in settings with very high bandwidth variability?
If not, how should congestion control
algorithms evolve? What is the effect of other delays in the core network,
and should the core network provide support such as split TCP or cross-layer
optimization? 

This paper aims at addressing such questions using a novel and very accurate end-to-end
simulation framework.  One of the significant challenges in 
evaluating the end-to-end performance in mmWave 
is the need to capture detailed aspects of the channel, link layer adaptation
and backhaul, all of which can have a significant impact at higher layers.
To this end, our simulation framework incorporates
complex measurement and ray-tracing 3D channel models with dynamics,
beamforming and tracking algorithms,
link-layer scheduling and retransmission mechanisms, and core network 
latency and bandwidth constraints. This level of detail results in more realistic scenarios, and therefore our framework can be used to study real-world problems.

The rest of the paper is organized as follows. In Section \ref{sec:e2e}, we describe the proposed end-to-end architecture. In Section \ref{sec:channel}, we describe the modeling of
 the channel dynamics of our mmWave channel, which is used to conduct a detailed performance evaluation, presented in Section \ref{sec:perfeva}. Finally, in Section \ref{sec:conclu}, we draw conclusions from the presented work.

\section{5G mmWave end-to-end framework}
\label{sec:e2e}
Building on the mmWave ns3 module presented in our earlier work
\cite{mezzavilla20155g}, we developed a 5G mmWave end-to-end simulation framework, described in \cite{russell2016ns3} and briefly summarized here. At the physical layer, the proposed module incorporates detailed modeling of the mmWave
channel that can capture spatial clusters, path dynamics,
antenna patterns and beamforming algorithms. The paths in the channel
can be modeled by either statistical clusters
\cite{rappaport2015wideband}, ray tracing~\cite{abdullah2015channel} 
or physical measurements\cite{rappaport2015wideband}.
Beamforming and channel tracking algorithms are also incorporated.

\begin{figure}[t!]
    \centering
    \includegraphics[trim={2.1cm 1.5cm 0 0},clip, width=0.5\textwidth]{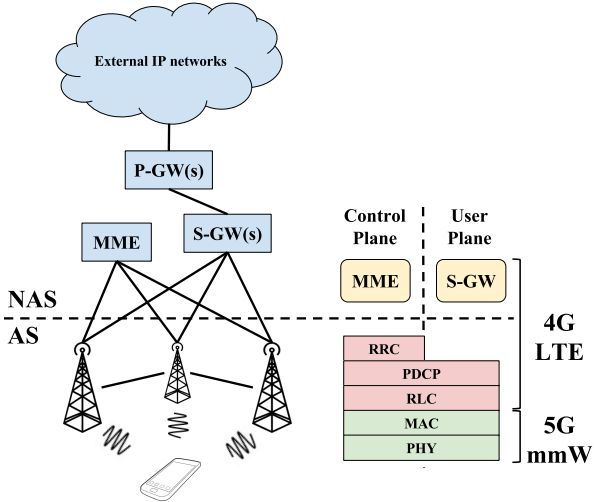}
    \caption{5G mmWave end-to-end framework. }
    \label{fig:e2e}
    \vspace{-5mm}
\end{figure}

The simulation framework also offers a highly configurable
MAC layer structure that can incorporate most of the recent frame designs including \cite{Dutta:15,khan2011millimeter}.
The design flexibility can be used to evaluate the effect of control,
scheduling and feedback delays. The radio link contol (RLC), radio resource control (RRC), non-access stratum (NAS) and core network
architecture are inherited from the original LENA module \cite{baldo2011ns}
to provide detailed modeling of all the higher layers of the 4G LTE stack. Fig.~\ref{fig:e2e}
 illustrates this architecture, which is composed of two main parts, namely, the \textbf{core network} and the \textbf{radio access network}.

The core network, also referred to as non-access stratum (NAS), comprises all the functions needed to establish the connection among cell towers and external IP networks. Each component is a separate server that executes a specific set of functions.

\textbf{P-GW:} The Packet Data Networks (PDN) Gateway enables the communication with external IP networks.

\textbf{S-GW:} The serving gateway is the mobility anchor
and acts as a router by forwarding data among base stations and PDN gateways.

\textbf{MME:} The mobility management entity controls high-level mobile operations through signaling and handover commands.

The radio access network, also referred to as access stratum (AS), 
is the component responsible for the wireless connection between a base station (BS) and a mobile device, called user equipment (UE) in 3GPP terminology.

\textbf{RRC:} This layer provides services and functions for broadcasting system information related to NAS and AS, paging, security, connection establishment, maintenance and release.

\textbf{PDCP:} This communication protocol is mainly used for header compression and decompression of IP data.

\textbf{RLC:} This layer operates in 3 modes: Transparent Mode (TM), Unacknowledged Mode (UM), and Acknowledged Mode (AM). Our module supports the UM and AM mode, including PDCP and RRC functionality on top of that. UM and AM mode offers RLC layer error correction, concatenation, segmentation and reassembly of RLC data units.

\textbf{MAC:} Our proposed mmWave MAC layer is designed to meet the ultra low latency and high data rate demands, as presented in \cite{Dutta:15}, thus following a flexible frame structure. A hybrid automatic repeat request (HARQ) is also implemented in order to better react to channel quality fluctuations.

\textbf{PHY:} We define here a set of mmWave-specific functions to capture transmission and reception capabilities trough a proper abstraction model, as presented in \cite{mezzavilla:MIESM}. The characterization of the propagation behavior in the new spectrum, which is fundamental for our framework, is reported in the next Section.

\section{MmWave channel dynamics}
\label{sec:channel}

\begin{figure*}[t!]
    \centering
    \begin{subfigure}[b]{0.49\textwidth}
        \includegraphics[width=\textwidth]{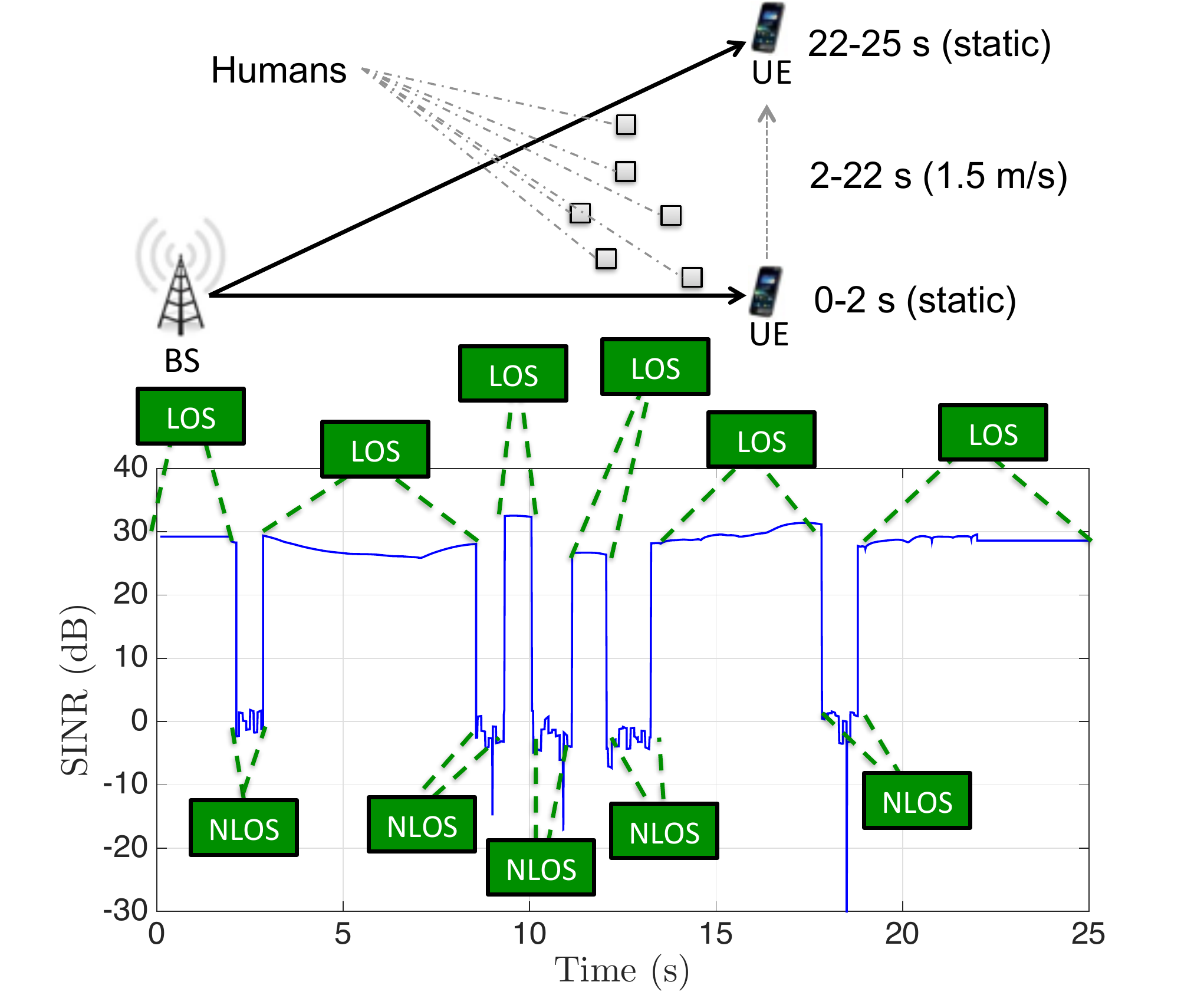}
        \caption{SINR vs time for Scenario $1$.}
        \label{fig:s1_sce}
    \end{subfigure}\quad
    \begin{subfigure}[b]{0.49\textwidth}
        \includegraphics[trim={10mm 1mm 25mm 1mm}, clip, width=\textwidth]{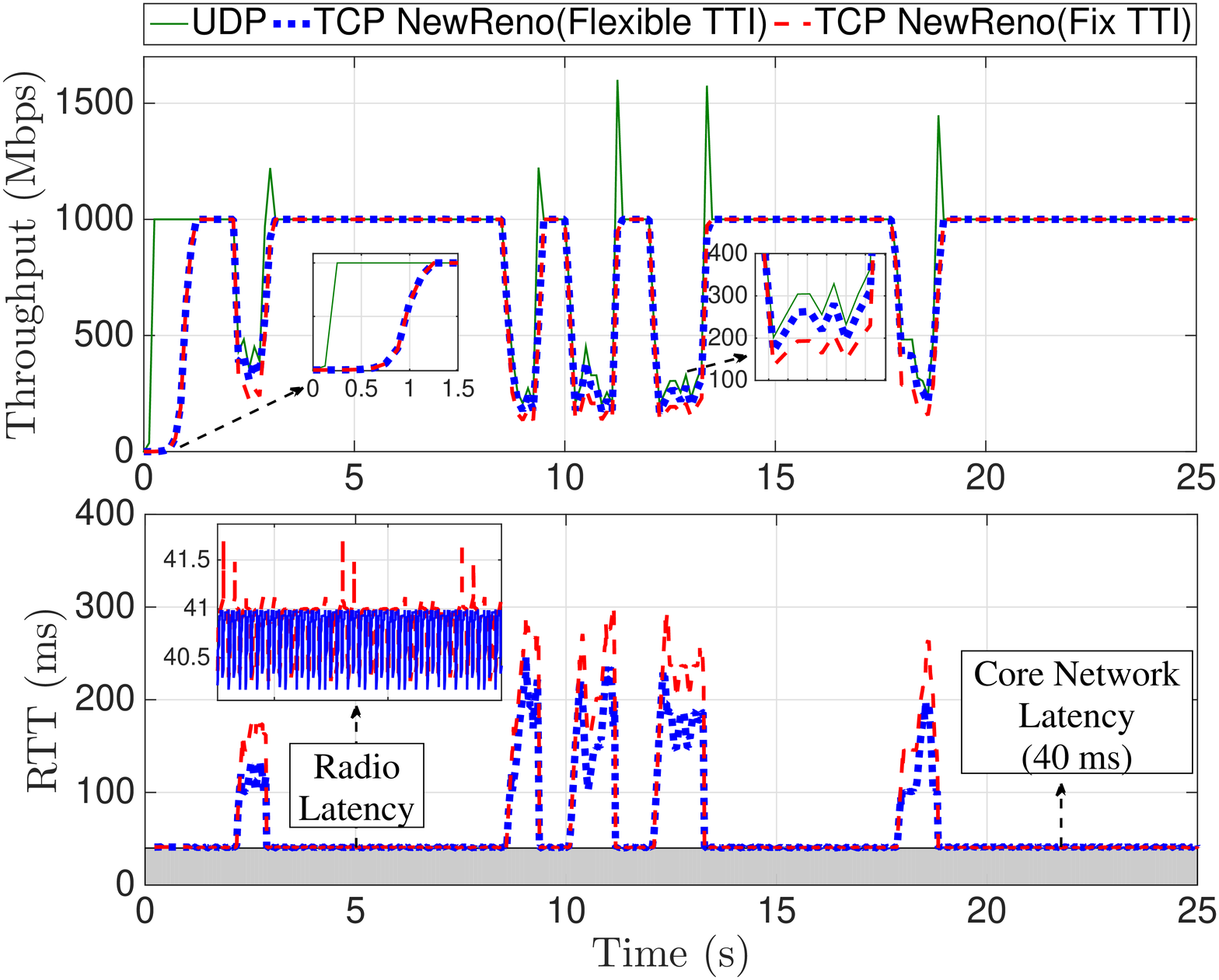}
        \caption{Performance comparison at $1$ Gbps data rate.}
        \label{fig:s1_tcp}
    \end{subfigure}\quad
    \caption{Simulation Scenario 1}\label{fig:scenario1}
    \label{fig:s1}
        \vspace{-5mm}
\end{figure*}

We aim at understanding how commonly used data applications will perform under \emph{massive capacity} and \emph{intermittency}. The massive capacity
derives from the large increase in the available bandwidth, possibly up to 
two orders of magnitude. The intermittency is caused by three factors, as mentioned in earlier comments: (i) dynamic multipath (things moving around us), (ii) changes in scattering as a result of our movement in the environment and (iii) small-scale fading as a result of diffuse scatter and  materials and surface types. At the present time, the dynamics of mmWave channels are not fully understood
and there is no single agreed upon statistical model for simulation
-- some very initial work can be found in \cite{eliasi2015stochastic,samimi2016mmWaveFading}.
The simulation framework proposed in \cite{russell2016ns3} is general, and in this work we consider
two possible models for channel intermittency: (i) a simplified
semi-statistical model with a
simulated topology and (ii) a more detailed ray tracing model.

\textbf{Semi-statistical model with simulated topology:} 
In this model, we first fix the location of the BS
transmitter, obstacles and the UE route.  
At each point along the route, we then determine if the UE is a in Line-of-Sight (LOS) or Non-Line-of-Sight (NLOS)
 state based on the existence of an unobstructed path.
In other words, if the line that connects the transmitter to the receiver crosses one or more buildings, the link is NLOS; otherwise, the link is LOS. Next, similar to 
\cite{mezzavilla20155g}, large scale parameters
of the channel (the path cluster powers, angles of arrival, angular spread, delay,
etc.) are drawn from the static statistical model proposed in \cite{rappaport2015wideband}.  
The large-scale parameters are then assumed to be constant over the entire route,
and we create a time-varying channel by adding the effect of the time-varying
distance path loss and of Doppler shifts on the paths (this will cause small scale fading
depending on the angular spread), and by eliminating the 
LOS paths in the regions where the UE is in a NLOS state.
Note that in this model, the LOS-NLOS transitions are immediate -- there is no modeling
of diffraction.

\textbf{Realistic traces:} As a second option,
we use raytracing to describe the time-correlated dynamics of the mmWave channel. To do so, we first generate the data of the sample points along a route measured in a real environment (Bristol, UK) and then apply raytracing to characterize the multipath. The ray model determines the mean power, time delay and 3D angles of the direct and scattered paths for a given city environment. Statistical small-scale variations (modeled based on measurements) are then added on a per-path basis to represent diffuse scattering from rough material surfaces. Finally, we feed the resulting traces into our simulator \cite{russell2016ns3}, where, as described in \cite{mezzavilla20155g,mattia_sharing}, we apply beamforming gains and channel tracking based on the channel characterization presented in \cite{samimi2016mmWaveFading,AkdenizCapacity:14}.
\section{Performance Evaluation}
\label{sec:perfeva}


In Section \ref{subsec:topo}, we aim at capturing the way the link transitions among the LOS, NLOS and Outage states affect the application layer in a standalone mmWave context. To do so, along with a mmWave BS and a mobile user, we drop different sized link-blocking cubes, and observe the end-to-end performance of the simulated scenarios. The carrier frequency is $28$ GHz, the transmit power $30$ dBm, the total bandwidth $1$ GHz, the number of TX and RX antennas is $64$ and $16$, respectively, and the core network latency is $20$ ms.

In Scenario $1$, we compare the performance of a single UDP flow against a TCP connection to observe (i) the time-offset experienced by TCP to reach capacity and (ii) the reduced TCP utilization due to the transmission of ACKs. Additionally, we will compare the TCP trends obtained with both a flexible and a fixed frame structure, to capture the gains achieved with the design proposed in \cite{Dutta:15}. Then, in Scenario $2$, we will compare the performance achieved by a TCP flow at varying data rates and RLC buffer sizes. Finally, in Scenario $3$, we will force some outage events to see how different variants of TCP (NewReno and Cubic) react to a retransmission timeout (RTO) timer expiration.

In Section \ref{subsec:real}, we investigate the behavior of TCP over spatially consistent channel traces that are obtained by applying ray tracing techniques (enhanced using small-scale mmWave measurements to capture the statistical elements of diffuse scatter) and validate the key findings captured through our simulations.


\begin{figure*}[t!]
    \centering
    \begin{subfigure}[b]{0.49\textwidth}
        \includegraphics[width=\textwidth]{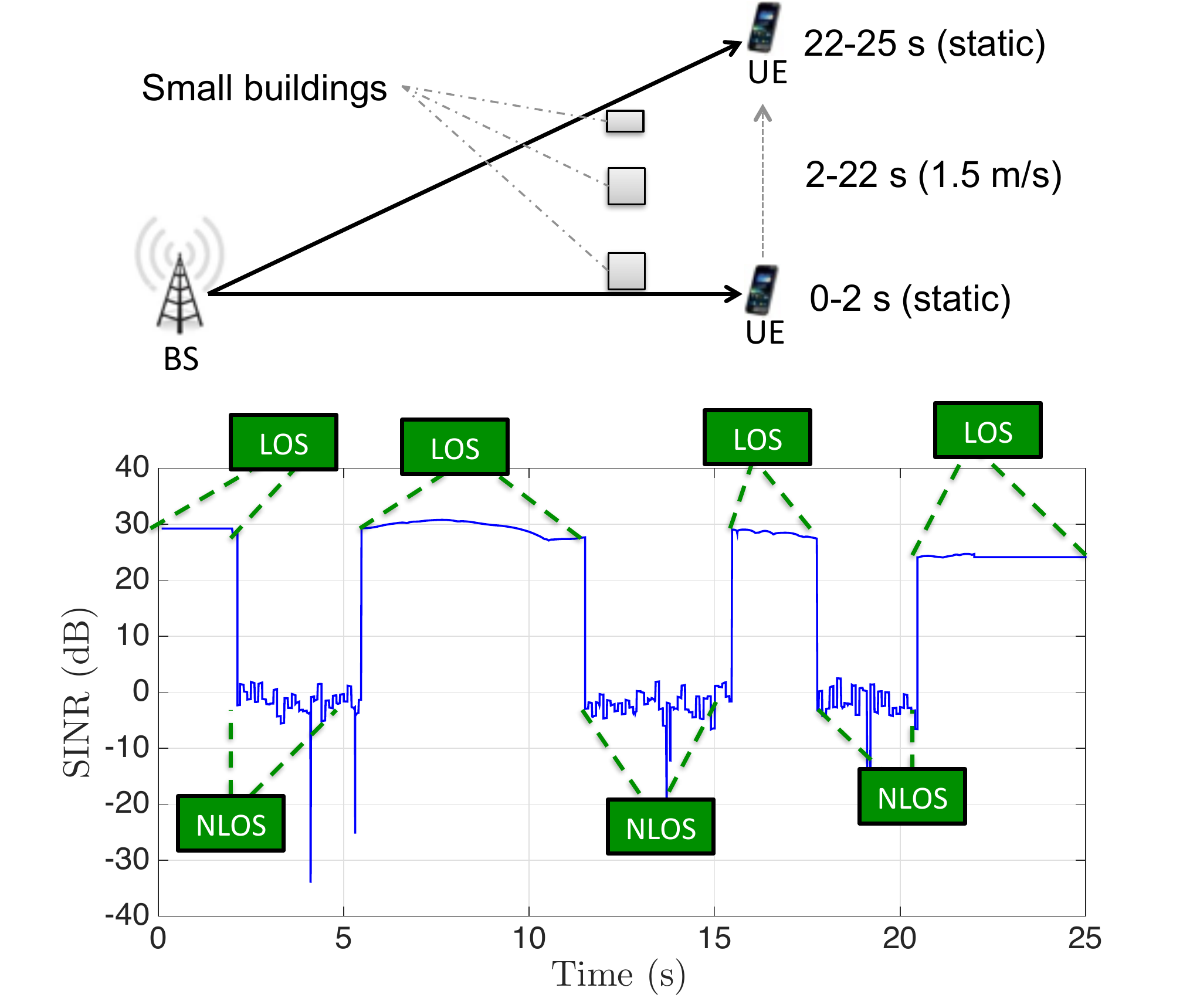}
        \caption{SINR vs time for Scenario $2$.}
        \label{fig:s2_sce}
    \end{subfigure}\quad
    \begin{subfigure}[b]{0.49\textwidth}
        \includegraphics[trim={10mm 1mm 25mm 1mm}, clip, width=\textwidth]{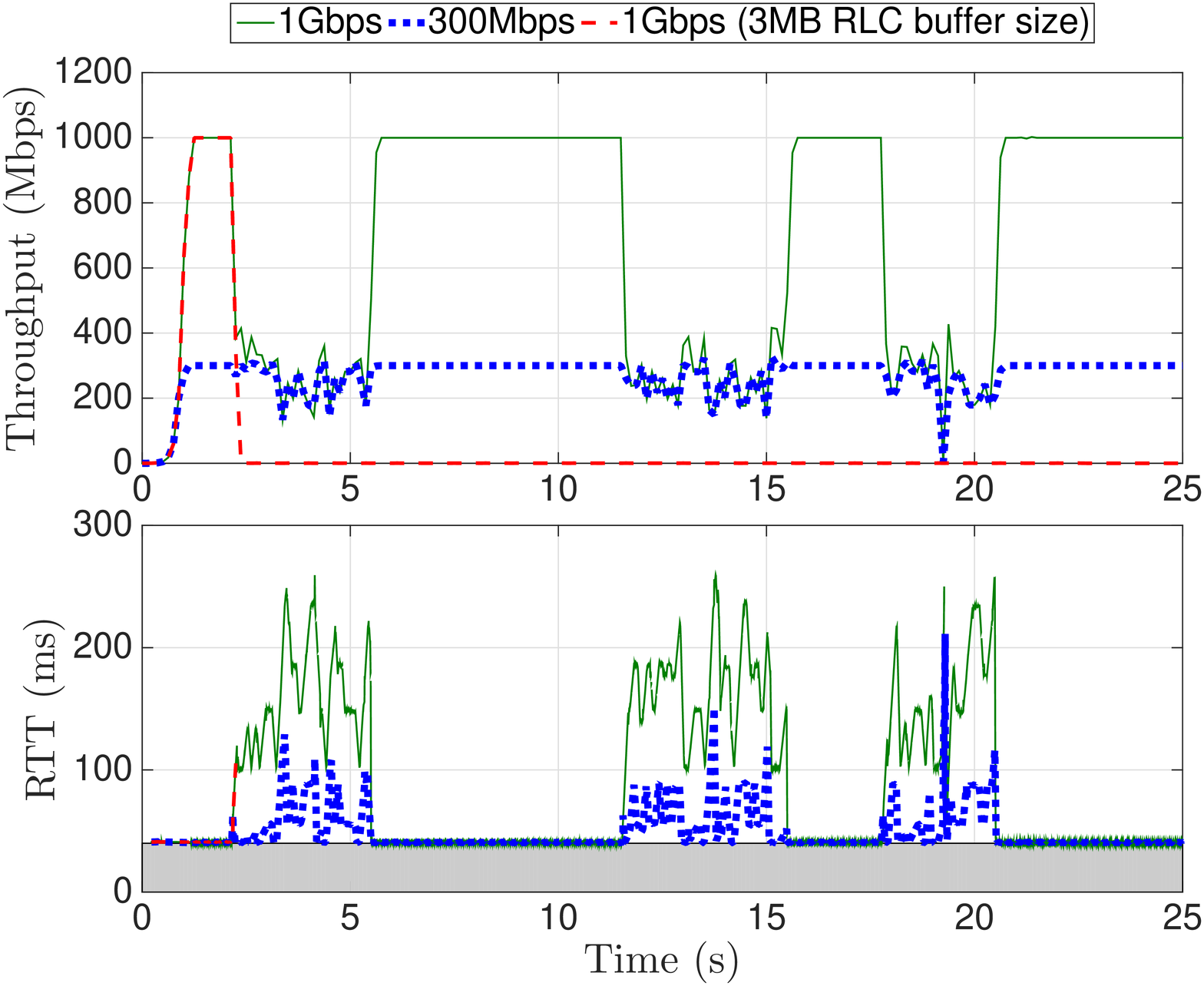}
        \caption{Performance comparison at $1$ Gbps data rate.}
        \label{fig:s2_tcp}
    \end{subfigure}\quad
    \caption{Simulation Scenario 2}\label{fig:scenario2}
        \vspace{-5mm}
\end{figure*}

\subsection{Simulated topology}
\label{subsec:topo}

\textbf{Scenario 1:} We let a UE move over a path characterized by sudden link transitions due to \emph{human} obstacles, represented as $0.5$ m $\times$ $0.5$ m cubes. TCP/UDP data packets are sent from a remote host to the UE with source rate $1$ Gbps. We report both the topology and the time evolution of the received power in Fig. \ref{fig:s1_sce}.

\emph{1) UDP vs. TCP -- } As expected, the throughput of the UDP flow ramps up to the source rate faster than TCP. This is due to the slow start phase that characterizes the TCP connection, which, as shown in Fig. \ref{fig:s1_tcp}, reaches the rate of  $1$ Gbps after approximately $1$ second. Additionally, we can observe how, in terms of achieved throughput, UDP slightly outperforms TCP in NLOS regimes because of the TCP ACKs transmitted in the uplink, that affect TCP downlink utilization. The UDP throughput spikes observed at each $NLOS \rightarrow LOS$ transition are due to the RLC buffer, which accumulates a number of packets during the reduced SINR regime, and transmits the entire burst when the high capacity of LOS condition, around $3.5$ Gbps as measured in \cite{russell2016ns3}, is restored (recall that no congestion control is used in UDP).

\emph{2) Flexible TTI vs. Fix TTI -- } As theoretically analyzed in \cite{Dutta:15}, we show here some numerical results for the throughput and latency achieved with a flexible frame structure. Due to a dynamic design, our flexible scheme shows a better utilization of the channel in both directions, uplink and downlink, resulting in higher throughput and lower latency, thus corroborating the validity of our proposed frame structure.

\emph{3) Latency -- } It is important to note how in our scenario, where the application server is far from the BS ($40$ ms round-trip latency), the bottleneck is the core network. In fact, thanks to a shorter TTI, the mmWave radio at LOS introduces only $1$ ms latency. On the other hand, in NLOS regimes, the latency increases by a factor of $100$. This is due to lower layer retransmissions and RLC buffer size, as better illustrated in the next scenario.

\textbf{Scenario 2:} In our second simulated scenario, the link-blocking cubes are buildings with different sizes. A UE moves along the same route as in the previous scenario, with the same pedestrian speed, i.e., $1.5$ m/s, while downloading a TCP NewReno flow. We report both the topology and the time evolution of the received power in Fig. \ref{fig:s2_sce}.

\emph{1) Bufferbloat -- } Bufferbloat, which causes high latency and packet delay variation when queueing too many packets in buffers, is a known problem \cite{gettys2011bufferbloat}. As also shown in the previous scenario, with NLOS connectivity the latency dramatically increases. Due to the tremendous capacity available in the mmWave bands, the TCP congestion window (CW) is very large. Thanks to lower layer retransmissions, the retransmission timeout (RTO) timer never expires because packets are never dropped. Hence, both the CW and the RLC buffer, which is set to $10$ MB by default, keep increasing. This directly translates into higher latency, thus introducing the problem of \emph{bufferbloat}. If we compare the solid green and dotted blue lines in Fig. \ref{fig:s2_tcp} that correspond to $1$ Gbps and $300$ Mbps data rates, respectively, we note that, while exhibiting a similar throughput, the latency is much higher at $1$ Gbps than at $300$ Mbps. The number of packets accumulated at the RLC buffer is indeed bigger, resulting in a more pronounced \emph{bufferbloat} effect.

\begin{figure*}[t!]
    \centering
    \begin{subfigure}[b]{0.49\textwidth}
        \includegraphics[width=\textwidth]{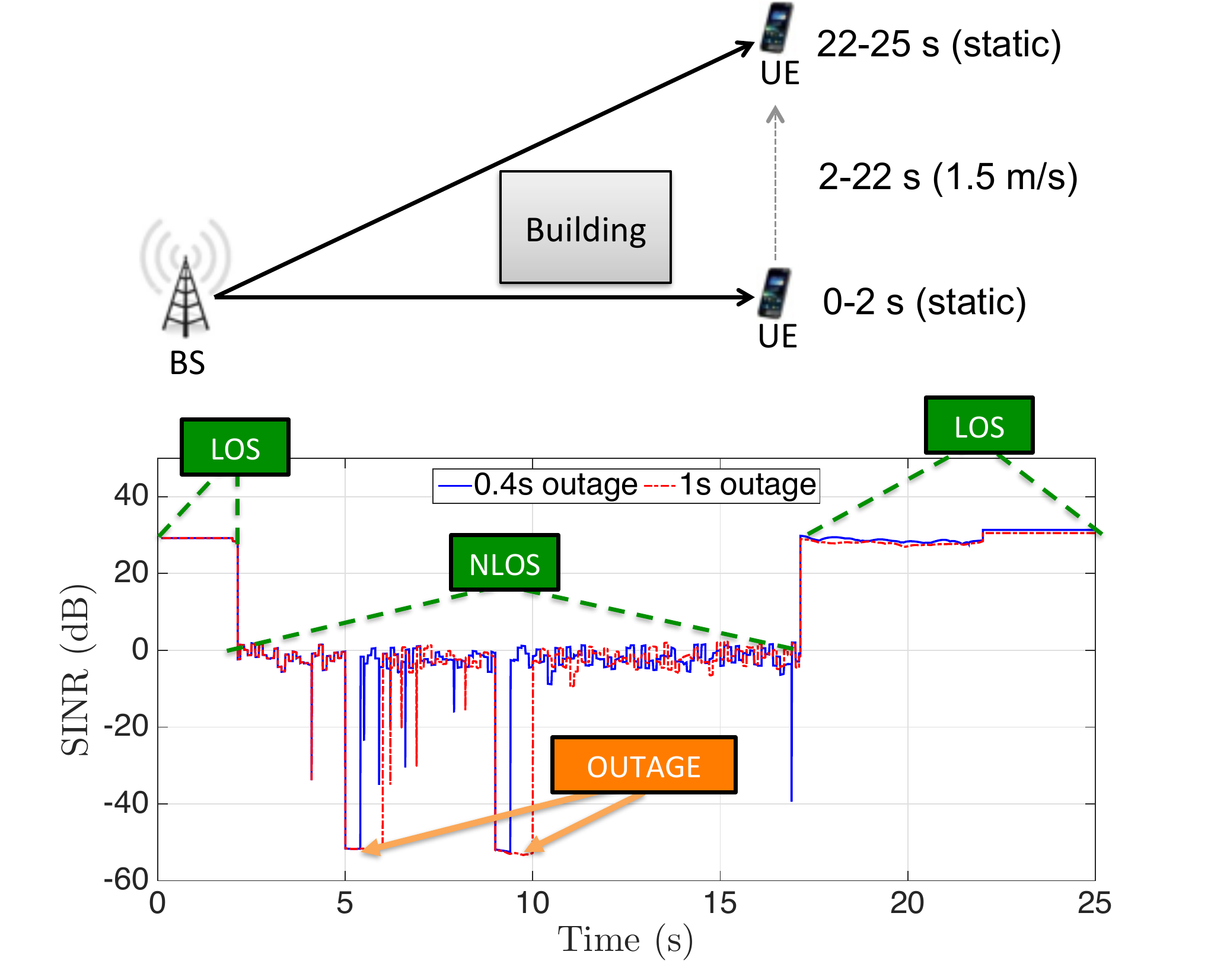}
        \caption{SINR vs time for Scenario $3$.}
        \label{fig:s3_sce}
    \end{subfigure}\quad
    \begin{subfigure}[b]{0.49\textwidth}
        \includegraphics[trim={10mm 1mm 25mm 1mm}, clip, width=\textwidth]{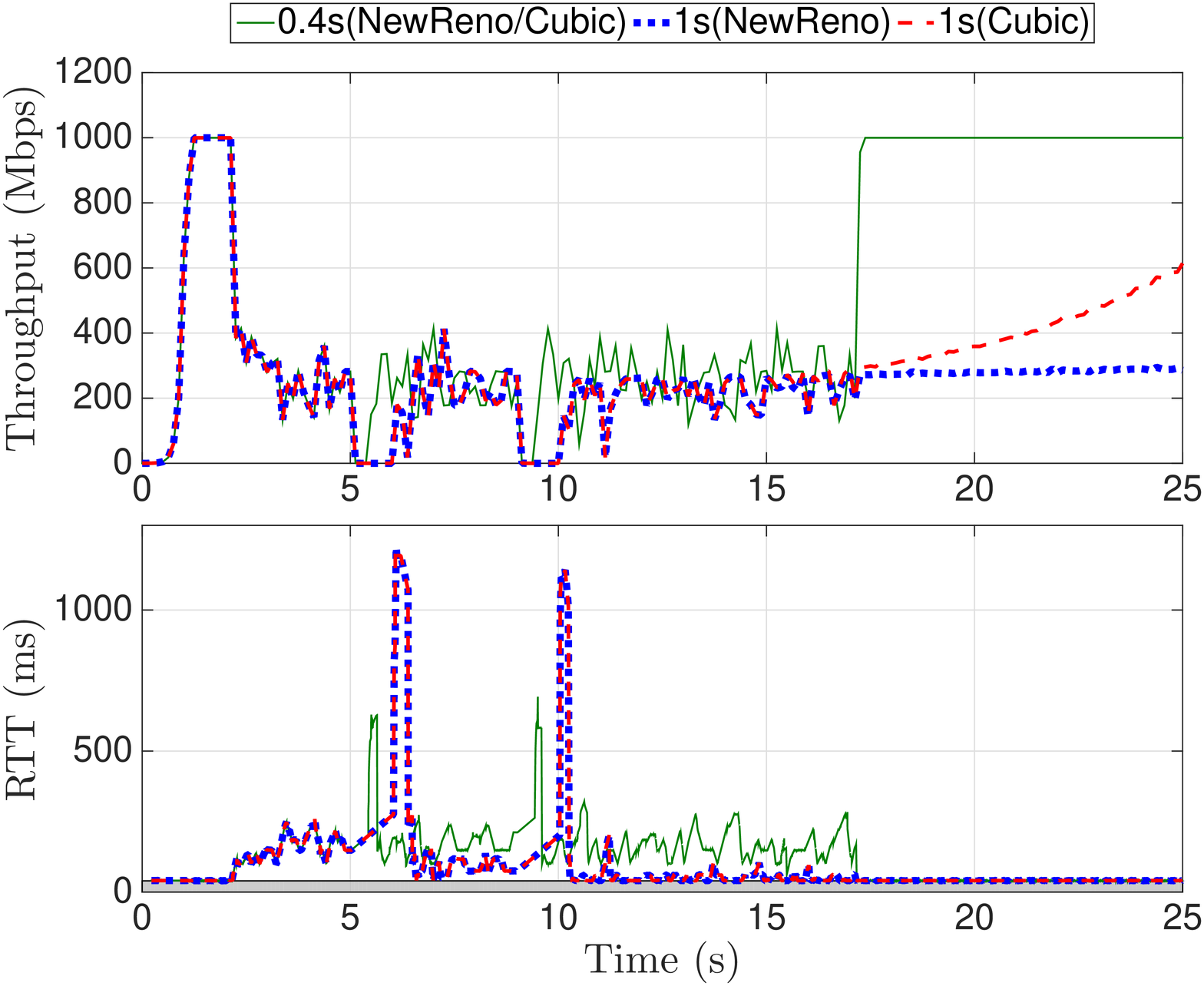}
        \caption{Performance comparison at $1$ Gbps data rate.}
        \label{fig:s3_tcp}
    \end{subfigure}\quad
    \caption{Simulation Scenario 3}\label{fig:scenario3}
        \vspace{-5mm}
\end{figure*}

\emph{2) Buffer overflow - Fast retransmit --} To overcome the \emph{bufferbloat} problem, we reduced the size of the RLC buffer to $3$ MB, as represented by the dashed line in Fig. \ref{fig:s2_tcp}. The results are catastrophic, as the buffer cannot keep up with the packet arrival rate enabled by the large TCP congestion window, and drops a large burst of packets which is proportional to the CW. Consequently, TCP enters a \emph{fast retransmit} phase, where $1$ packet is retransmitted every RTT, thus requiring, in this case where the RTT is $40$ ms, almost $40$ seconds to retransmit $1000$ dropped packets. The reason why  TCP timeouts never occur is that the NewReno implementation in ns3 version 3.24 resets the retransmit timer after each partial acknowledgement -- this is called the Slow-but-Steady variant of TCP NewReno as mentioned in \cite{parvez2006tcp}.

\textbf{Scenario 3:} In the above scenarios, no TCP retransmissions are triggered thanks to lower layer retransmission schemes. Nonetheless, we would like to observe how different TCP variants, namely NewReno and Cubic, react to link failures that can cause an RTO expiration. To do so, we forced two outage events, of length $0.4$ and $1$ s, respectively. Topology and received power are reported in Fig. \ref{fig:s3_sce}.

\emph{1) Short outage --} For the $0.4$ s outage case, with RLC acknowledge mode enabled, HARQ at the MAC layer and ARQ at the RLC layer can still recover the corrupted packets without triggering any TCP retransmission. As a consequence, the CW never drops and the connection reaches capacity almost instantaneously, as shown in Fig. \ref{fig:s3_tcp}.

\emph{2) Long outage -- } On the other hand, when the outage is $1$ s long, the TCP RTO expires. After a timeout event, both TCP variants, Cubic and NewReno, enter slow start by setting the CW to $1$ and halving the slow start threshold. Due to small CW, the latency is reduced, but this has a dramatic impact on the throughput of both TCP implementations, as shown in Fig. \ref{fig:s3_tcp}. In particular, even though Cubic has been designed for better efficiency through a faster CW growth, this feature is only able to cope with some impairments and turns out to be unable to work properly if the outages are too long.  NewReno is even worse since it has linear CW growth during the congestion avoidance phase. Therefore, even this highly optimized TCP version may not work well in our environment if outages occur on a certain time scale, which we were able to characterize. We believe that these observations will stimulate further research towards designing novel congestion control procedures able to better adapt to the unique mmWave radio environment.

\begin{figure*}[t!]
    \centering
    \begin{subfigure}[b]{0.275\textwidth}
        \includegraphics[width=\textwidth]{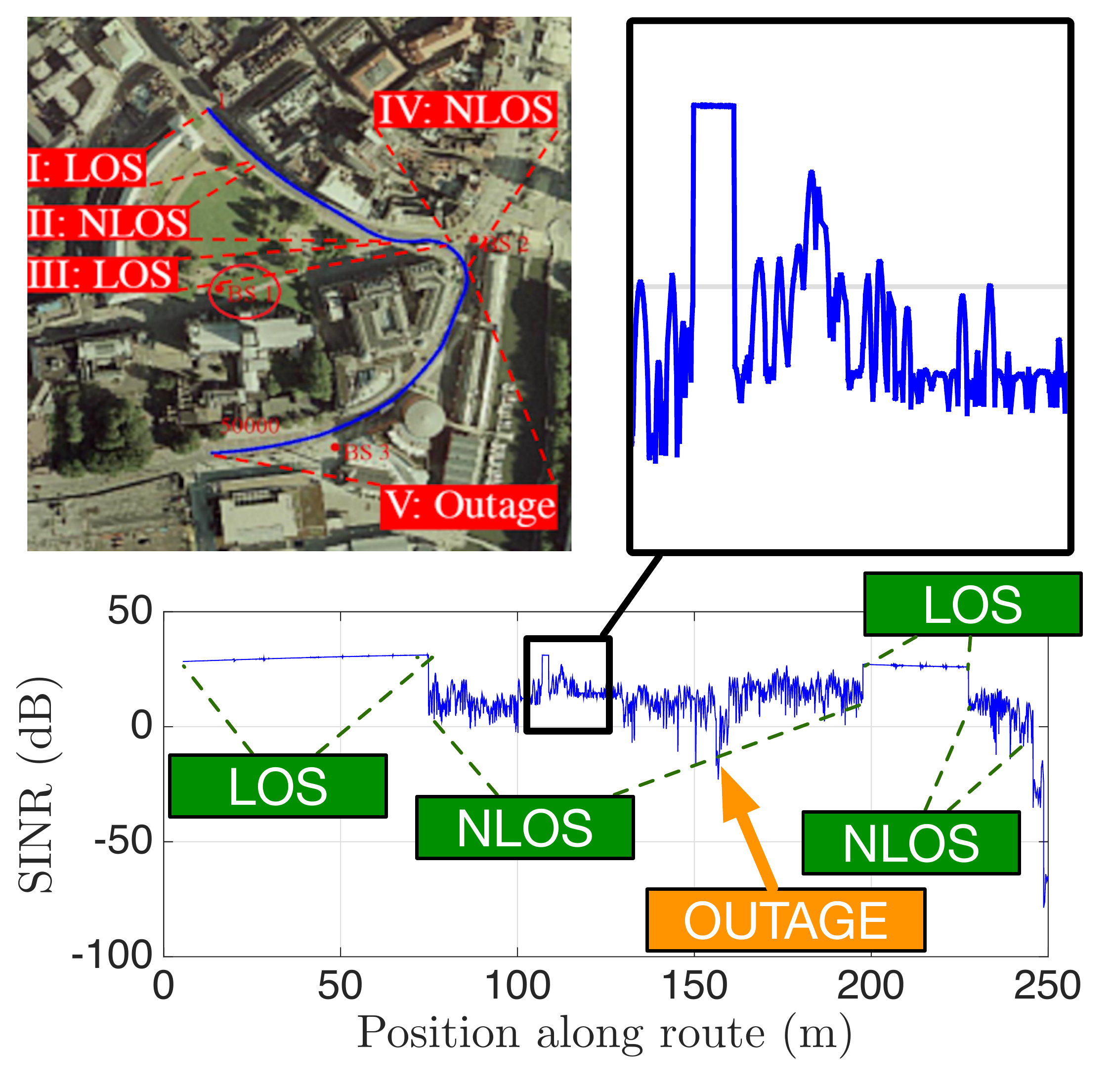}
        \caption{SINR}
        \label{fig:raytra_sce}
    \end{subfigure}\quad
        \begin{subfigure}[b]{0.3425\textwidth}
        \includegraphics[trim={2mm 1mm 25mm 0mm}, clip,width=\textwidth]{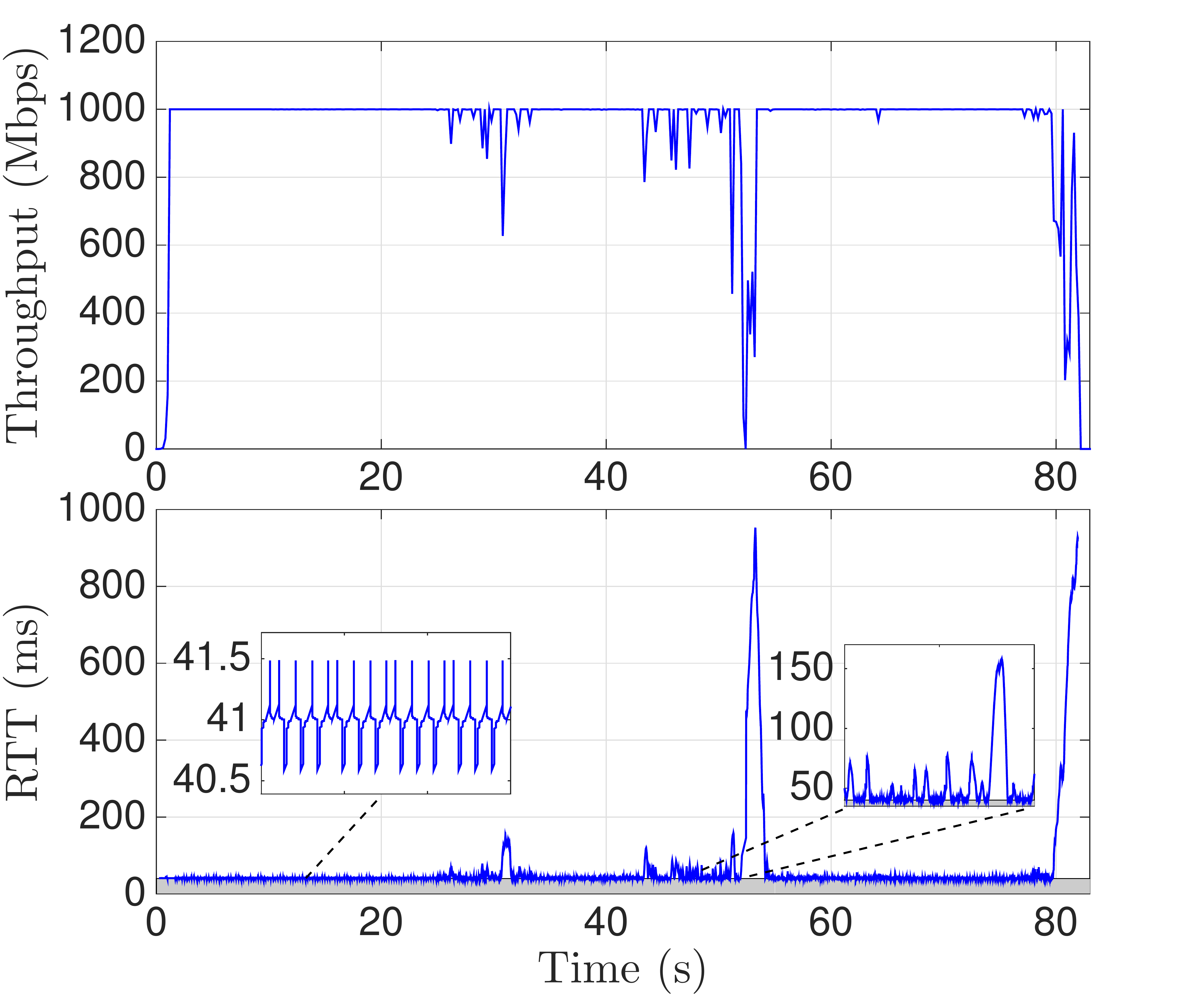}
        \caption{TCP performance (3 m/s)}
        \label{fig:raytra_tcp2}
    \end{subfigure}\quad
        \begin{subfigure}[b]{0.3425\textwidth}
        \includegraphics[trim={2mm 1mm 25mm 0mm}, clip,width=\textwidth]{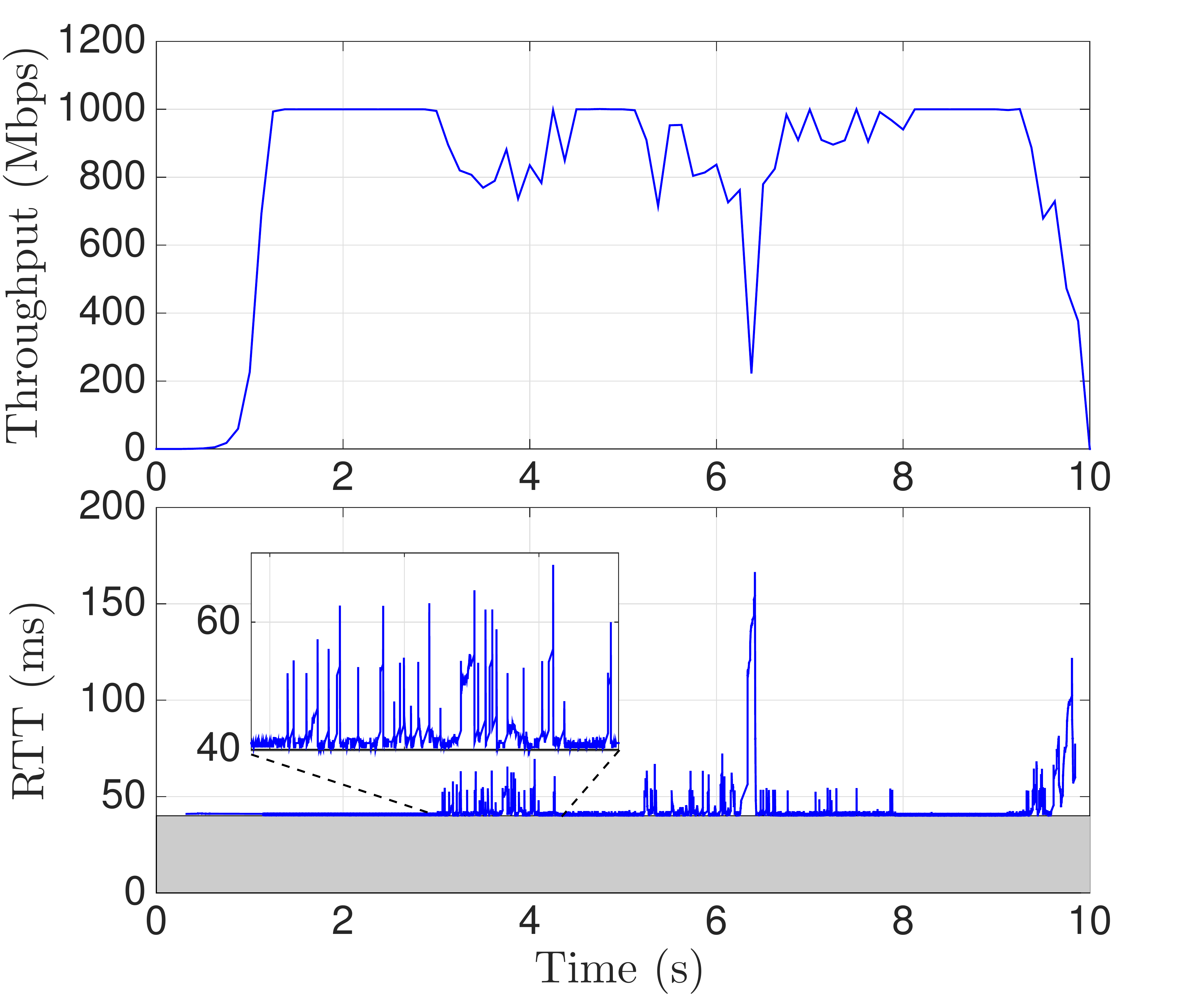}
        \caption{TCP performance (25 m/s)}
        \label{fig:raytra_tcp1}
    \end{subfigure}\quad
    \caption{Raytracing Route}\label{fig:scenario3}
\end{figure*}

\subsection{Realistic traces}
\label{subsec:real}

For a more realistic analysis, we plugged a trace generated through a combination of diffuse scatter measurements and raytracing into our end-to-end mmWave simulation framework, and compared the TCP rate and latency at (pedestrian) $3$ m/s and (vehicular) $25$ m/s UE mobility, respectively.

As shown in Fig. \ref{fig:raytra_sce}, we observe some $LOS \leftrightarrow NLOS$ transitions that are determined by a group of trees across the direct path between the UE and the transmitting BS. The SINR dynamics is similar to those obtained through simulations in our earlier scenarios., i.e., $\sim30$ dB. As noted before, and as presented in Figs. \ref{fig:raytra_tcp2} and \ref{fig:raytra_tcp1}, these sudden channel drops do not trigger any TCP retransmission. Hence, the only negative trend is the increased latency experienced during the NLOS regimes, which relates to the \emph{bufferbloat} problem described above. Additionally, we can note in Fig. \ref{fig:raytra_sce} that the link breaks ($NLOS \rightarrow Outage$) at approximately $150$ m from the beginning of the \emph{raytraced} route. Because the RTT never exceeds the RTO timer\footnote{The default minimum RTO is 1 s.}, RLC and MAC retransmission schemes prevent any TCP retransmission which, as shown above, would have dramatically affected the TCP performance.

Finally, if we compare Figs. \ref{fig:raytra_tcp2} and \ref{fig:raytra_tcp1}, we can observe that, at the higher UE speed, the achieved throughput is lower and the latency spike generated during the outage event is lower. The former is explained by the fact that the CQI feedback is not frequent enough, given the rapid fluctuations of the channel due to the high speed, which results in some performance degradation. The latter, instead, is simply related to the fact that the fast-moving user is in outage for a much shorter amount of time.

\section{Conclusions}
\label{sec:conclu}
MmWave cellular systems are likely to have very high peak capacity,
that however is also highly variable.  The basic question in this paper was to understand
how this variability impacts the end-to-end performance.
This evaluation is challenging since the end-to-end performance depends in complex ways on
the interactions of the underlying channel dynamics, beam tracking,
MAC-layer scheduling and retransmissions, network delays and congestion control.
To understand these system-wide implications, 
we have studied some representative scenarios using a novel end-to-end ns3-based simulation framework.Our initial evaluations have revealed several possible problems in 
existing congestion control protocols in mmWave links.
First, due to the very high data rate, current slow start mechanisms can take
several seconds to achieve the full throughput offered by a mmWave PHY-layer.
This may be problematic in applications that rely on short TCP connections.
Second, large drops in rate, which are likely to be common in LOS-NLOS
transitions, can result in very high levels of queuing and buffering,
dramatically increasing the latency. Thirdly, after a retransmission timeout (RTO),
even aggressive TCP protocols such as Cubic can take inordinately long to recover
to full rates.  While RLC and MAC layer retransmissions can shield upper layers
from packet losses, RTO will still occur in real situations.

It is important to further study the mmWave channels and understand the effects of propagation dynamics on the upper layers. Additionally, how networks should evolve to address these problems and realize the full potential 
of the mmWave bands remains a broad and open problem.  
Solutions could involve bringing content closer to the edge
(thus reducing delay), assistance within the core network, or changes in the congestion
control protocols themselves.

\bibliographystyle{IEEEtran}
\bibliography{biblio}

\end{document}